\documentclass[conference]{IEEEtran}
\usepackage{url}
\usepackage{verbatim} % allows multiline comments
\usepackage[tight,footnotesize]{subfigure} % for the subfloat in the figs
\usepackage{multirow}
\usepackage{algorithm}
\usepackage[noend]{algorithmic}
\usepackage{xspace}
\usepackage{graphicx}
\usepackage{epsfig}
\usepackage{amsmath}
\usepackage{amssymb}
\usepackage{Definitions}
\usepackage{empheq}
\usepackage{multirow,amscd,amssymb}
\usepackage{colortbl}
\usepackage{slashbox,lscape}
\usepackage{paralist}

\newcommand{\hide}[1]{}

\newcommand{\explain}[2]{\underset{\mathclap{\overset{\uparrow}{#2}}}{#1}}
\newcommand{\explainup}[2]{\overset{\mathclap{\underset{\downarrow}{#2}}}{#1}}

\graphicspath{{./FIG/}}

\begin{document}

\title{Modeling Adoption and Usage of Competing Products} %in Social Media}

\author{\IEEEauthorblockN{Isabel Valera}
\IEEEauthorblockA{MPI for Software Systems\\
ivalera@mpi-sws.org}
\and
\IEEEauthorblockN{Manuel Gomez-Rodriguez}
\IEEEauthorblockA{MPI for Software Systems\\
manuelgr@mpi-sws.org}}
%\and
%\IEEEauthorblockN{Krishna Gummadi}
%\IEEEauthorblockA{MPI for Software Systems\\
%gummadi@mpi-sws.org}}

%\author{\IEEEauthorblockN{Name}
%\IEEEauthorblockA{Affiliation\\
%name@domain}
%\and
%\IEEEauthorblockN{Name}
%\IEEEauthorblockA{Affiliation\\
%name@domain}
%\and
%\IEEEauthorblockN{Name}
%\IEEEauthorblockA{Affiliation\\
%name@domain}}

\maketitle

\begin{abstract}
The emergence and wide-spread use of online social networks has led to a dramatic increase on the availability of social activity data. 
Importantly, this data can be exploited to investigate, at a microscopic level, some of the problems that have captured the attention of economists, 
marketers and sociologists for decades, such as, \eg, product adoption, usage and competition. 

In this paper, we propose a continuous-time probabilistic model, based on temporal point processes, for the adoption and frequency of use of competing products, where the frequency 
of use of one product can be modulated by those of others.
This model allows us to efficiently simulate the adoption and recurrent usages of competing pro\-ducts, and generate traces in which we can easily recognize the 
effect of social influence, recency and competition.
We then develop an inference method to efficiently fit the model parameters by solving a convex program. The problem decouples into a collection of 
smaller subproblems, thus scaling easily to networks with hundred of thousands of nodes. 
We validate our model over synthetic and real diffusion data gathered from Twitter, and show that the proposed model does not only provides a good 
fit to the data and more accurate predictions than alternatives but also provides interpretable model parameters, which allow us to gain insights into 
some of the factors driving product adoption and frequency of use.

%We propose a data-driven model, based on the continuous-time Hawkes process, for the adoption of competing products and conventions in social networks. The proposed model naturally models cooperation and competition among products by incorporating interpretable parameters that help us to understand the competing dynamics in social networks. We then propose an inference method that efficiently fits the model parameters by solving a set of convex optimization problems and that handles networks with hundred of thousands of nodes. We validate the performance of the proposed method over synthetic and real diffusion data. For the experiments over real data, we use the data gathered from Twitter from March 2006 to September September 2009, in which we study two types of competing contagions: URL shortening services and retweet conventions. We show that the proposed method not only presents good predictive power but also allows us to investigate the competing dynamics in the network.

\end{abstract}

\section{Introduction}
\label{sec:intro}
There is a long history of work in economics and mar\-ke\-ting on studying product adoption, product competition and product life cycle~\cite{mahajan1990new}. 
This work has typically developed mathematical models, such as, the Bass diffusion model~\cite{bass1969new}, the probit model~\cite{mcfadden1980econometric}, 
or models based on information cascades~\cite{bikhchandani1992theory}, that attempt to capture the macroscopic evolution of a product or set of competing products 
but not its (their) microscopic evolution, \ie, product adoptions and recurrent usage by specific individuals.
This has been in part due to the lack of large-scale fine-grained product adoption and product usage data, which would allow to propose and validate microscopic models
at scale.

The emergence and wide-spread use of online social networks has led to a dramatic increase on the availability of large-scale fine-grained social activity data, in 
which all individual adoptions and subsequent uses of a product are often \emph{observable}.
This availability has opened up a great opportunity for a paradigm shift, where we investigate both the macroscopic and microscopic dynamics of product adoption and 
usage.
As a consequence, there recent empirical studies have investigated the adoption and usage of competing products~\cite{antoniades2011we} % and social conventions~\cite{kooti2012emergence} 
in online social networks.
Therefore, it is necessary to develop microscopic models of product adoption and usage, which are largely non-existent to date.

\subsection{Our approach}
We introduce a continuous-time probabilistic model, based on temporal point processes, for the adoption and frequency of use of competing products in online social networks. The model captures several intuitive key factors,
which has been previously discussed in the literature:
\begin{itemize}
% \denselist
%
\item[I.] {\bf Social Influence.} Whenever a user observes that one or more of her neighbors are using a product, she may decide to start using the product~\cite{kempe03maximizing}.

\item[II.] {\bf Recency.} If a user has used a product in the past, she is more likely to use the product in the future than if she has never used it~\cite{anderson2014dynamics}.

\item[III.] {\bf Competition.} If a user who is currently using a pro\-duct $p_1$ observes that one or more of her neighbors use a competing product $p_2$, she may decide to switch from
product $p_1$ to $p_2$~\cite{bharathi2007competitive}.
\end{itemize}

To simultaneously capture the above mentioned factors, which have been typically studied separately, our model leverages multivariate Hawkes processes~\cite{Liniger2009}. Although Haw\-kes processes 
have been used for modeling positive social influence before~\cite{blundell2012modelling,farajtabar2014activity,iwata2013discovering,ZhoZhaSon13,ZhoZhaSon13b}, the key innovation here is that we 
explicitly model recency and competition, and allow for both positive and negative social influence. As a consequence, our design is especially well fitted to model adoption and recurrent usage of competing products.

Perhaps surprisingly, the flexibility of our model of competing products does not prevent us from efficiently simulate from the model, and learn the model parameters from real world product fine-grained usage 
data:
\begin{itemize}
% \denselist
%
\item {\bf Efficient simulation.} We develop an efficient sampling method that leverages Ogata'{}s method~\cite{ogata1981lewis}, and scales linearly in the number of products. Our method exploits the sparsity of 
social networks to reduce the overall complexity (refer to Algorithm~\ref{alg:simulation}).

\item {\bf Efficient parameter estimation.} We fit the parameters of the model using historical data by solving a maximum likelihood estimation problem, which reduces to solving a convex program. 
Importantly, the problem decouples in several smaller pro\-blems, allowing for natural parallelization so that we can fit the parameters of the model for networks with hundred of thousands of nodes.
\end{itemize}

Moreover, we validate our model using both synthetic and real social activity data, and show that: 
\begin{enumerate}
% \denselist
%
\item Our model can generate product use events that obey temporal patterns described in related literature~\cite{barabasi05human, shukla2004effect}, such as 
temporal bursts of events or product switching.

\item Our regularized maximum likelihood estimation method can recover the parameters of the model with an accuracy that increases as we record more product 
uses over time.

\item Our model provides a significantly better fit and more accurate product use event predictions in real data gathered from 
Twitter~\cite{cha2010measuring} than several alternatives~\cite{anderson2014dynamics, barabasi05human}.
In particular, we experiment with one type of competing product, url shortening services~\cite{antoniades2011we}, and one type of conventions, the way Twitter users indicated a tweet was a retweet back in 2009~\cite{kooti2012emergence}.

\item Our model allows us to gain insights into some of the factors driving product adoption and frequency of use.
For example, we find that the usage of popular products is typically driven by recency, however, an exposure to a less popular product can have a strong inhi\-biting effect 
on future uses of popular products.
\end{enumerate}

\subsection{Related work}
Several models of adoption of competing products in online social networks have been proposed in the computer science literature in recent years~\cite{bharathi2007competitive, dubey2006competing, goyal2012competitive}.
In particular, Bharathi et al.~\cite{bharathi2007competitive} extend the independent cascade model, Dubey et al.~\cite{dubey2006competing} propose a local quasilinear model, and Goyal et 
al.~\cite{goyal2012competitive} consider a broad class of local influence processes.
Moreover, competition has been also studied in other contexts: meme diffusion~\cite{myers12clash}, rumor propagation~\cite{kostka2008word}, spread of misinformation~\cite{budak2011limiting}, 
opinion formation~\cite{degroot1974reaching,friedkin1990social} and disease spread~\cite{beutel2012interacting,prakash2012winner}.
However, most previous work share the following limitations, which we tackle in our work.
First, they consider discrete-time sequential models of competition, which are difficult to estimate accurately from real world data~\cite{lesong2012nips,manuel11icml,iwata2013discovering}.
Instead, we model competition using\- a continuous-time asynchronous model, designed to naturally fit the event data we record (\ie, the times in which users use products).
Second, they consider each user to adopt only one \emph{contagion} (be it a product, a meme, rumor, opinion or disease), the one she adopted \emph{first}, and \emph{use} it only once. In contrast, 
we allow users to use it once or several times and adopt different competing contagions simultaneously. That is, we model users'{} frequency of use, and how this \emph{frequency} depends on what
users \emph{observe} over time.
Hence, we are able to model non-progressive phenomena~\cite{kempe03maximizing}. 
Third, they only allow for pairwise interactions between contagions, being able to study competition between only two different products. In contrast, we consider n-ary interactions, which consider the interactions/competition among all the competing products in the network. 
 
Repeat online behavior have been studied in the specific contexts of repeat queries in web search~\cite{teevan2006history,teevan2007information} or repeat website visits~\cite{adar2008large,adar2009resonance}. 
Only very recently, Anderson et al.~\cite{anderson2014dynamics} propose a fairly general discrete-time model of online reconsumption of goods, such as songs or movies, and experiences. 
However, this line of work does not model social influence nor competition and typically focus on goods, experiences or behaviors a user may get satiated or bored of, rather than products a user 
may repeatedly keep using for long periods of time.

Last, Hawkes processes have been also applied to a large variety of problems in which self-excitement plays a fundamental role: earth quake prediction~\cite{marsan2008extending}, 
crime prediction~\cite{egesdal2010statistical}, computational neuro\-science~\cite{krumin2010correlation}, or high frequency trading~\cite{bacry2013modelling}.
However, to the best of our knowledge, Hawkes processes have not been applied in the context of adoption and usage of competing products.

\section{Proposed model}
\label{sec:formulation}
?fiIn this section, we formulate our probabilistic model of adoption and frequency of use of competing products. We start by revisiting the foundations of temporal point processes, and then 
describe the specifics of our model.

\subsection{Temporal Point Processes}
A temporal point process is a random process whose rea\-li\-za\-tion consists of a list of discrete events localized in time, $\cbr{t_i}$ with $t_i \in \RR^+$ and $i \in \ZZ^+$. Many different types 
of data produced in online social networks and the Web can be represented as temporal point processes, \eg, the times when a user uses a product. 
A temporal point process can be equivalently re\-pre\-sen\-ted as a counting process, $N(t)$, which records the number of events before time $t$, \eg, the number of product use events before time $t$. 
Let the history $\Hcal(t)$ be the list of times of events $\cbr{t_1, t_2, \ldots, t_n}$ up to but not in\-clu\-ding time $t$. Then, in a small time window $dt$ between $[0, t)$, the number of observed event is 
\begin{align}
  \label{eq:count_change}
  d N(t) = \sum_{t_i \in \Hcal(t)} \delta(t-t_i) \, dt,  
\end{align}
where $\delta(t)$ is a Dirac delta function, and hence $N(t) = \int_0^t dN(s)$. It is often assumed that only one event can happen in a small window of size $dt$, and hence $dN(t) \in \cbr{0,1}$. 

An important way to characterize temporal point processes is via the conditional intensity function --- the stochastic model for the time of the next event given all the times of previous events. 
The conditional intensity function $\lambda^*(t)$ (intensity, for short) is the conditional probability of observing an event in a small window $[t, t+dt)$ given the history $\Hcal(t)$, \ie,
\begin{align}
  \label{eq:intensity}
  \lambda^*(t)dt = \PP\cbr{\text{event in $[t, t+dt)$}|\Hcal(t)}, 
\end{align}
where $^*$ means that the function $\lambda^*(t)$ may depend on the history $\Hcal(t)$. Based on the intensity, one can obtain the conditional expectation of the number of events in the windows 
$[t,t+dt)$ and $[0,t)$, respectively, as
\begin{align}
  \EE[dN(t) | \Hcal(t)] &= \lambda^*(t)\, dt,~\text{and} \\
  \EE[N(t)|\Hcal(t)] &= \int_0^t \lambda^*(\tau)\, d \tau
\end{align}
The functional form of the intensity $\lambda^*(t)$ is often designed to capture the phenomena of interests~\cite{aalen2008survival}. For example, the following forms have been argued for in the 
information propagation and social influence literature~\cite{du13nips,farajtabar2014activity,manuel11icml,survival13icml}:
\begin{itemize}
% \denselist
  \item {\bf Poisson process.} The intensity is assumed to be independent of the history $\Hcal(t)$ and constant,~\ie, 
  \begin{equation} \label{eq:poisson}
  \lambda^*(t) = \lambda,
  \end{equation} 
  where $\lambda \geqslant 0$ is a scale parameter.
  \item {\bf Weibull renewal process.} The intensity is assumed to be dependent only on the last event $t_n \in \Hcal(t)$ before $t$,~\ie, 
  \begin{equation} \label{eq:weibull}
  \lambda^*(t) = k (t - t_n)^{k-1} \lambda^{k} \geqslant 0, 
  \end{equation}
  where $k>0$ is a shape parameter and $\lambda>0$ is a scale parameter. Remarkably, if the shape parameter is less than $1$, then the inter-event time distribution is heavy-tailed, as observed in real social activity data~\cite{barabasi05human}.
  \item {\bf Hawkes Process.} The intensity captures a mutual excitation phenomena between events,~\ie,
  \begin{align}
    \label{eq:hawkes}
    \lambda^*(t) = \mu + \alpha \sum_{t_i \in \Hcal(t)} g_{\omega}(t-t_i),
  \end{align}
  where $g_{\omega}(t)$ is a nonnegative triggering kernel such that $g_{\omega}(t) = 0$ for $t < 0$, $\mu\geqslant 0$ is a baseline intensity independent of the history, and the summation of kernel terms is history dependent and 
  a stochastic process by itself. Here, the occurrence of each historical event increases the intensity by a certain amount $\alpha \geqslant 0$. 
  %This summa\-tion term can also be written in convolution form  
  %\begin{align}
  %  \hspace{-3mm}
  %  \label{eq:kernel_convolve} 
  %  \sum_{t_i \in \Hcal(t)}g_{\omega}(t-t_i) 
  %  = \underbrace{\int_{0}^t g_{\omega}(t-\tau)\, dN(\tau)}_{:=g_{\omega}(t)\, \star\, dN(t)}. 
  %\end{align}
\end{itemize}
Now, given a time $t' \geqslant t$, we can also characterize the conditional probability that no event happens during $[t, t')$ and the conditional probability density that an event occurs at time $t'$ using\- 
the intensity $\lambda^*(t)$ by means of the following well-known relationships~\cite{aalen2008survival}:
\begin{align}
  S^*(t') &= \exp\rbr{-\int_t^{t'} \lambda^*(\tau) \, d\tau},~~\text{and}~~ \label{eq:survival_fun}\\ 
  f^*(t') &= \lambda^*(t')\, S^*(t').\label{eq:density_fun}
\end{align}
With these two quantities, we can express the log-likelihood of a list of event times $\cbr{t_1,t_2,\ldots,t_n}$ in an observation window $[0, T)$ with $T\geqslant t_n$ as 
\begin{align}
  \label{eq:loglikehood_fun}
  \Lfra = \sum_{i=1}^n \log \lambda^*(t_i) - \int_{0}^T \lambda^*(\tau)\, d\tau,  
\end{align}
which will be useful for learning the parameters of our model from observed data. With the above background in temporal point processes, we can now proceed to detail the formulation of the proposed model for 
product adoption and frequency of use, including our choice of specific functional form for the intensity $\lambda^{*}(t)$.

\subsection{Model for Product Adoption} \label{sec:model-adoption}
Given a social network $\Gcal = (\Vcal, \Ecal)$ and a set of (related) competing products $\Pcal$, we model the generation of product use events by users in the network using temporal point
processes. Here, we record each event as a triple
\begin{align}
  \label{eq:event}
%   \boxed{
  e ~~:=~~ (~\explain{u}{\text{user}},~~~\explainup{p}{\text{product}},~~~\explain{t}{\text{time}}~),
%   } 
\end{align}
where the triplet means that the user $u \in \Vcal$ used product $p \in \Pcal$ at time $t$. We assume that we can observe the triple $e$ for 
every product use and there are not attribution problems.
Given a list of use events $\{e_1 = (u_1, p_1, t_1), \ldots, e_n = (u_n, p_n, t_n)\}$ up to time $t$, the history $\Hcal_u(t)$ of product use events by 
user $u$ is
\begin{equation}
\Hcal_u(t) = \{e_i = (u_i, p_i, t_i) | u_i = u\ \mbox{and}\, t_i<t\},
\end{equation}
and the history $\Hcal_{up}(t)$ of use events of product $p$ by user $u$ is
\begin{equation}
\Hcal_{up}(t) = \{e_i = (u_i, p_i, t_i) | u_i = u,  p_i = p\ \mbox{and}\, t_i<t \}.
\end{equation}
Finally, the entire history of product use events is denoted as $\Hcal(t) := \cup_{u \in \Vcal} \Hcal_u(t)$.
Note that each node $u$ does not necessarily use all products $\Pcal$ and can use a specific product $p$ once or several times.

Now, we use a set of counting processes, one for each user and product, to record the generated product use events. 
More specifically, we denote the set of counting processes as a matrix $\Nb(t)$ of size $|\Gcal| \times |\Pcal|$ for each 
fixed time point $t$. The $(u, p)$-th entry in the matrix, $N_{up}(t) \in \cbr{0} \cup \ZZ^+$, counts the number of times user $u$ 
used product $p$ up to time $t$. That is, $N_{up}(t)=|\Hcal_{up}(t)|$ is the size of the history $\Hcal_{up}(t)$. 
Then, we can characterize the users'{} product usage over time using their corresponding intensities as
\begin{equation}
\EE[d\Nb(t)\, |\,  \Hcal(t)] = \Lambdab^*(t) \, dt,
\end{equation}
where the matrix $d\Nb(t):=\rbr{~dN_{up}(t)~}_{u \in \Vcal, p \in \Pcal}$ contains the number of product use events per user and product in the window $[t, t+dt)$, and the matrix $\Lambdab^*(t) := (~\lambda_{up}^*(t)~)_{u \in \Vcal, p \in \Pcal}$ 
contains the intensities associated to all the user-product pairs at time $t$. The sign $^{*}$ means that the intensity matrix $\Lambda^*(t)$ depends on the history $\Hcal(t)$ and each particular intensity $\lambda_{up}^{*}(t)$ depends on the history node $u$ is exposed 
to, $\cup_{v \in \Ncal(u) \cup \{u\}} \Hcal_v(t)$, where $\Ncal(u) = \{ v \in \Vcal | (v, u) \in \Ecal \}$ denotes the nodes $u$ follows (\ie, her neighbors). 
In the next section, we specify the functional form of the intensities that accounts for both recency and social influence.
\begin{figure*}[t]
\centering
\subfigure[Burstiness]{\includegraphics[width=.38\textwidth]{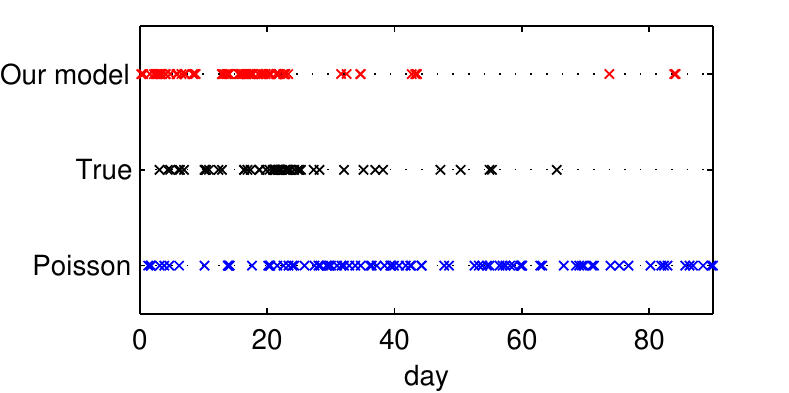}\label{fig:bursts}} \hspace{5mm}
\subfigure[Product switch]{\includegraphics[width=.38\textwidth]{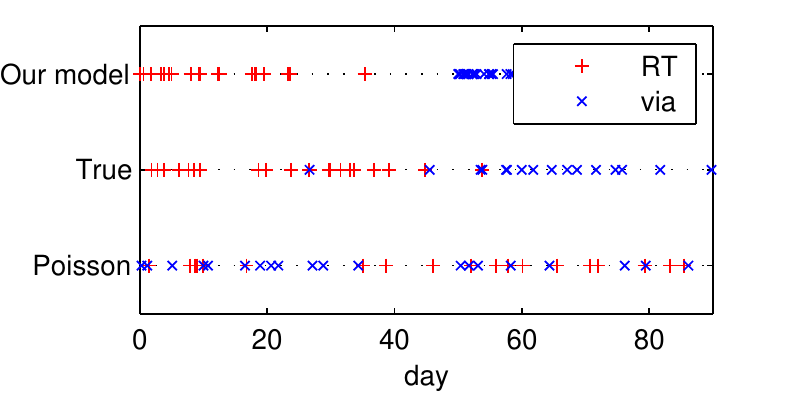}\label{fig:competition}}
\caption{Example of event burstiness and product competition. Our model is able to generate bursts and product switches due to considering mutual excitation and inhibition between events in the 
intensities, given by Eq.~\ref{eq:hazard-general}, while Poisson processes fail to do so. Since product uses are realizations of a (stochastic) temporal point process, one cannot expect to precisely 
match the timing of each individual event.}% True number of events is 87; the Hawkes model predicts 89 events; and the Poisson model predicts 88 events.}
\label{fig:burst}
\end{figure*}

\subsection{Intensity for Product Adoption}\label{sec:intensities}
To model the intensities, $\Lambda^*(t)$, for product use events, we leverage the multivariate version~\cite{Liniger2009} of the Hawkes process defined by Eq.~\ref{eq:hawkes}. However, in our work, we modify the 
original definition of Hawkes processes to allow for both excitation and inhibition phenomena between product use events~\cite{ogata1998linear}. 
In particular, each intensity $\lambda_{up}^*(t)$ takes the following form: 
\begin{equation}\label{eq:hazard-general}
\begin{split}
\lambda_{up}^*(t) & := \underbrace{\mu_p}_{\text{spontaneous use}} + \quad \underbrace{\sum_{l=1}^{|\Pcal|} a_{lp} \sum_{e_i \in \Hcal_{ul}(t)} g_{\omega}(t - t_i)}_{\text{recency}} \\ 
& \quad + \underbrace{\sum_{l=1}^{|\Pcal|} b_{lp} \sum_{v \in \Ncal(u)} \sum_{e_i \in \Hcal_{vl}(t)} g_{\omega}(t - t_i)}_{\text{social influence}},
\end{split}
\end{equation}

The first term $\mu_p \geqslant 0$ denotes the base event intensity, which models the spontaneous adoption of a product by a user on her own initiative.
In other words, it models adoptions due to other factors different than recency or social influence.
The second term models recency, where $a_{lp} \in \RR$ denotes the influence that a previous use of a product $l$ by the user has on her intensity for 
product $p$.
The third term models the propagation of peer influence over the network. Here, $b_{lp} \in \RR$ denotes the influence that a previous use of
a product $l$ by one of her neighbors has on the user'{}s intensity for product $p$.
Intuitively, one may expect that $a_{pp}, b_{pp} \geqslant 0$, since previous uses of a product may increase the probability that a user uses the same 
product in the future, while $a_{lp}, b_{lp} \leqslant 0$, $l \neq p$, since previous uses of a product may decrease the probability that a user use a competing 
product in the future. 
For conciseness, we stack all the model parameters in a vector and two matrices, \ie, $\mub := (\mu)_{p \in \Pcal}$, $\Ab = (a_{lp})_{l, p \in \Pcal}$ and $\Bb = (b_{lp})_{l, p \in \Pcal}$.
Finally, the triggering kernel $g_{\omega}(t)$ models the decay of influence over time. Here, for simplicity, we opt for an exponential kernel $g_{\omega}(t) := \exp(-\omega t)\II[t\geqslant 0]$.
However, since our inference method does not depend on this particular choice, one may decide to use more complicated kernels with some desirable characteristics~\cite{ZhoZhaSon13b} or perform a non-parametric estimation 
of the kernels~\cite{bacry2012non}. 
Remarkably, using exponential kernels allow for slightly more efficient likelihood evaluations and, as a result, more efficient model simulation and estimation, as discussed
in Sections~\ref{sec:simulation}~and~\ref{sec:estimation}.
%
% For example, a power law kernel $g_{\omega}(t) := \omega t^{-1-\omega}\II[t\geqslant 0]$ may be able to reproduce event data the lack of characteristic time scale.
%
%For conciseness, we stack all the model parameters in a vector and two matrices, \ie, $\mub := (\mu)_{p \in \Pcal}$, $\Ab = (a_{lp})_{l, p \in \Pcal}$ and $\Bb = (b_{lp})_{l, p \in \Pcal}$.

%It is remarkable and we can then express the intensities associated to a user $u$, $\lambdab_{u}^*(t) = (\lambda_{up}^*(t))_{p \in \Pcal}$, very compactly as:
%
%\begin{equation}\label{eq:hazard-general-matrix}
%\lambdab_{u}^*(t) = \mub + \Ab \left(g_{\omega}(t) \star d\Nb_{u}(t) \right) + \Bb \sum_{v \in \Ncal(u)} \left(g_{\omega}(t) \star d\Nb_{v}(t)\right),
%\end{equation}
%
%where $\mub = (\mu_p)_{p \in \Pcal}$, $\Ab = (a_{lp})_{l, p \in \Pcal}$,  $\Bb = (b_{lp})_{l, p \in \Pcal}$ and $\Nb_{v}(t) = (N_{vp})_{p \in \Pcal}$.

% Unless otherwise specified, we will consider throughout the rest of the paper the more general framework provided by the model in (\ref{eq:hazard-general}). 
%

\section{Model Simulation} \label{sec:simulation}
In this section, we first describe an efficient sampling me\-thod for our model of product adoption and then elaborate on the temporal patterns described in related literature~\cite{shukla2004effect} that our model is able 
to reproduce.

\subsection{Simulation method} 
We simulate samples (product usages) from our model by leveraging Ogata'{}s thinning algorithm~\cite{ogata1981lewis}. However, an off-the-shelf implementation of Ogata'{}s algorithm 
would need $O(n^2 |\Vcal|^2)$ operations to draw $n$ samples. Here, we first exploit the sparsity of social networks:
whenever we sample a new product use event from the model, only a small number of intensity functions, in the local neighborhood of the node that used the product, changes. As a 
consequence, we do not have to re-evaluate all intensity functions.
Moreover, for exponential trigger kernels, each time we re-evaluate an intensity function, we exploit the properties of the exponential function to do so in $O(1)$: let $t_1$ and $t_2$ be two 
consecutive events, then, we can compute each intensity $\lambda_{up}^*(t_2)$  as $(\lambda_{up}^*(t_1) - \mu_p) \exp(-\omega (t_2-t_1)) + \mu_p$.

The complete simulation algorithm is summarized in Algorithm~\ref{alg:simulation}, which needs $O(n d |\Vcal|)$ operations to draw $n$ samples, where $d$ is the maximum number of neighbors per
node.
Note that, since the model parameters $(a_{lp})_{l, p \in \Pcal}$ and $(b_{lp})_{l, p \in \Pcal}$ may take negative values, it may happen that $\lambda_{up}^*(t) < 0$ for some time $t$. In those 
cases, as proposed by Ogata~\cite{ogata1998linear}, we set $\lambda_{up}^*(t)$ to zero.

\subsection{Properties of the simulated product adoption}
By modeling recency and social influence, our model is able to generate \emph{realistic} collections of product use events, which obey several temporal patterns observed in real social activity 
data~\cite{barabasi05human, shukla2004effect}. Here, we pay attention to two of these temporal patterns, which alternative models based on Poisson processes or Weibull renewal processes
fail to capture:
\begin{itemize}
% \denselist
%
\item {\bf Bursts.} Social events are characterized by bursts of rapidly occurring events separated by long periods of inactivity~\cite{barabasi05human}. In our Twitter dataset (Refer to Section~\ref{sec:real-experiments}), we 
also find such bursts of events, for example, Figure~\ref{fig:bursts} shows the times (in black) in which a user used url shortening services (\eg, `bit.ly'). Remarkably, our model (in red) is able to mimic 
such bursts of events while alternative models, such as Poisson processes (in blue), fail to do so. Here, we trained the parameters of both models using the user'{}s real events (Refer to Section~\ref{sec:estimation}). 
Our model is able to generate bursts due to considering mutual excitation and inhibition between events in the intensities, given by Eq.~\ref{eq:hazard-general}.

\item {\bf Product Switches.} A person who uses a product may decide to switch to a competing product due to multiple factors, including social influence~\cite{shukla2004effect}. In our Twitter dataset (Refer to 
Section~\ref{sec:real-experiments}), we often find users that decide to switch products, for example, Figure~\ref{fig:competition} shows the times (in the center) in which a user switch from one retweet convention, `RT' (in red), to
another, `via' (in blue). Our model (in the top) is able to generate such a pro\-duct switch while a Poisson process (in the bottom) does not succeed. Again, we trained the parameters of both models using the user'{}s real events 
(Refer to Section~\ref{sec:estimation}).
Our model is able to generate product switches because it models competition by considering inhibition between events in the intensities, given by Eq.~\ref{eq:hazard-general}.
\end{itemize}
\begin{algorithm}[t]                      % enter the algorithm environment
\caption{Efficient Model Simulation}          % give the algorithm a caption
\label{alg:simulation}                           % and a label for \ref{} commands later in the document
\begin{algorithmic}[1]                    % enter the algorithmic environment
    \REQUIRE $n^{up}=0$ for ${u=1,\ldots,|\Vcal|, p=1,\ldots,|\Pcal|}$
    \STATE $I^* \leftarrow I^{|\Vcal|\times |\Pcal|}(t_0)\leftarrow \sum_{u}^{|\Vcal|} \sum_{p}^{|\Pcal|} \lambda_p^u (t_0)$\\
\hspace*{-19pt}\textbf{ Generate first event}:\\ 
    	\STATE Generate $q \sim \Ucal_{[0,1]}$ and $s \leftarrow t_0-\frac{1}{I^*} \ln (q)$\\
    	\STATE \textbf{if} $s>T$, \textbf{then} go to last step.	\\
    	\STATE \textbf{else} \textbf{Attribution Test:} \\
    	 \quad i) Sample $d \sim \Ucal_{[0,1]}$ \\
    	 \quad ii) Choose $u$ and $p$ such that $\frac{ I^{up-1}(t_0) }{I^*} < d \leq \frac{I^{up}(t_0)}{I^*}$ \\
    	 \quad iii) Set $t_1 \leftarrow t_1^{up}\leftarrow s$, $i \leftarrow 1$ and $n^{up} \leftarrow 1$
    \ENSURE 
    \WHILE{$s<T$}
       \STATE $I^* \leftarrow I^{|\Vcal|\times |\Pcal|}(t_{i}) + \sum_{pl}^{|\Pcal|} a_{pl}^u + \sum_{u' \in \Ncal(u)} b_{pl}^{u'} $
        \STATE Generate $q \sim \Ucal_{[0,1]}$
        \STATE Update $s \leftarrow t_i-\frac{1}{I^*} \ln (q)$
    	\STATE \textbf{if} $s>T$, \textbf{then} go to last step	
    	\STATE \textbf{else} \textbf{Attribution-Rejection Test:} \\
    	 \quad i) Sample $d \sim \Ucal_{[0,1]}$ \\
    	 \quad ii) \textbf{if} $d\leq \frac{I^{|\Vcal|\times |\Pcal|}(s)}{I^*}$, \textbf{then}\\
    	   \quad \quad - Choose $u$ and $p$ such that\\
    	    \quad \quad \quad \quad$\frac{ I^{up-1}(t_0) }{I^*} < d \leq \frac{I^{up}(t_0)}{I^*}$ \\
    	  \quad \quad - Set $t_{i+1} \leftarrow t_{n^{up}+1}^{up}\leftarrow s$, $i \leftarrow i+1$ \\ \quad \quad \quad and $n^{up} \leftarrow n^{up} +1 $\\
    	 \quad iii) \textbf{else} \\
    	  \quad \quad - Update $I^* \leftarrow I^{|\Vcal|\times |\Pcal|}(s)$ and go to step 8.
    \ENDWHILE{\textbf{end while}}
    \STATE \textbf{Output:} Retrieve the simulated process\\ $(\{t_{i}^{up}\} )_{u=1,\ldots,|\Vcal|, p=1,\ldots,|\Pcal|}$ on $[t_0, T ]$
\end{algorithmic}
\end{algorithm}

We remark that here our goal here is to show that our model can generate product use events obeying several temporal patterns observed in real social activity . However, since product uses are realizations of a (stochastic) 
continuous-time temporal point process, one cannot expect to precisely match the timing of each individual event.

\section{Model Parameter Estimation} \label{sec:estimation}
%
%Perhaps surprisingly, we can efficiently learn the parameters of our model using a set of historical product use events. 
In this section, we show how to efficiently learn the parameters of our model using a set of historical product use events.  
Given a collection of product use events $\Hcal(T) = \{ (u_i, p_i, t_i) \}$ recorded during a time period $[ 0, T)$ in a
social network $\Gcal = (\Vcal, \Ecal)$, our goal is to find the optimal parameters $\mub$, $\Ab$ and $\Bb$ by solving a regularized maximum likelihood estimation (MLE) problem.
To this end, we first compute the log-likelihood of the recorded events, using Eq.~\ref{eq:loglikehood_fun}, as follows:
\begin{equation}\label{eq:user-loglikelihood}
\mathcal{L}(\mub,\Ab, \Bb) = \sum_{e_i \in \Hcal(T)} \log \lambda_{u_i p_i}^*(t_i) - \sum_{u \in \Gcal} \sum_{p \in \Pcal} \int_{0}^T \lambda_{u p}^*(\tau)\, d\tau.
\end{equation}
where we can further express the integral terms as:
\begin{equation}\label{eq:user-loglikelihood-2}
\begin{split}
\int_{0}^T \lambda_{up}^*(\tau)\, d\tau &= T \mu_p  + \sum_{l=1}^{|\Pcal|} a_{lp} \sum_{e_i \in \Hcal_{ul}(t)} G_{\omega}(T - t_i) \\
& + \sum_{v\in \Ncal(u)} \sum_{l=1}^{|\Pcal|} b_{lp} \sum_{e_i \in \Hcal_{vl}(t)} G_{\omega}(T - t_i),
\end{split}
\end{equation}
where $G_{\omega}(t) = \int_{0}^{t} g_{\omega}(t') dt'$.

Then, we can formulate the regularized MLE problem as:
\begin{equation}
\label{eq:opt-problem}
\begin{aligned}
& \underset{ \mub, \Ab, \Bb}{\text{minimize}}
& & -\mathcal{L}(\mub, \Ab, \Bb) +\beta\, \Rcal(\mub, \Ab, \Bb), \\
& \text{subject to}
& &\mu_p \geq 0,
\end{aligned}
\end{equation}
where the first term is the negative log-likelihood of the events and the second term is the regularization term, being $\beta$ the parameter that controls the trade-off between these two terms. 
As long as we choose a convex regularizer, it is easy to show that this regularized 
MLE problem is jointly convex in $\boldsymbol{\mu}$, $\mathbf{A}$ and $\mathbf{B}$ by using linearity, composition rules for convexity, and the concavity property of the lo\-ga\-rithm. Hence, the global optimum can be 
found by many well-known algorithms~\cite{ZhoZhaSon13,ZhoZhaSon13b}. In practice, we solved Eq.~\ref{eq:opt-problem} with \texttt{CVX}, a software package for specifying and solving convex programs~\cite{cvx},
and used a quadratic re\-gu\-la\-rizer, $\Rcal(\mub, \Ab, \Bb) = \|\mub\|_2^2 + \|\Ab\|_2^2 + \|\Bb\|_2^2$, which we found to work well in practice.

For simplicity, we have described the model using the same model parameters, $\mub$, $\Ab$ and $\Bb$, for all users. However, in our experiments, we use different model parameters, $\mub^{u}$, $\Ab^{u}$ and $\Bb^{u}$, 
for each user. By doing so, we can decompose the model estimation procedure into $|\Vcal| \times |\Pcal|$ independent maximum likelihood estimation problems, one per user $u$ and product $p$. 
We can solve each subproblem in parallel, obtaining local solutions that are globally optimal.  Additionally, under exponential kernels, we can precompute all sums of triggering kernels and integrals of triggering kernels for each subproblem in linear
time, \ie, $O(|\Hcal_u(T)| + |\cup_{v \in \Ncal(u)} \Hcal_v(T)|)$, by exploiting the pro\-per\-ties of the exponential function, similarly as described in Section~\ref{sec:simulation}.
As a result, our model estimation procedure scales to networks on the order of hundreds of thousands of nodes.

% \section{Experiments}
%
\section{Experiments on synthetic data} \label{sec:synthetic-experiments}
\begin{figure}[t]
\centering
\includegraphics[width=.42\textwidth]{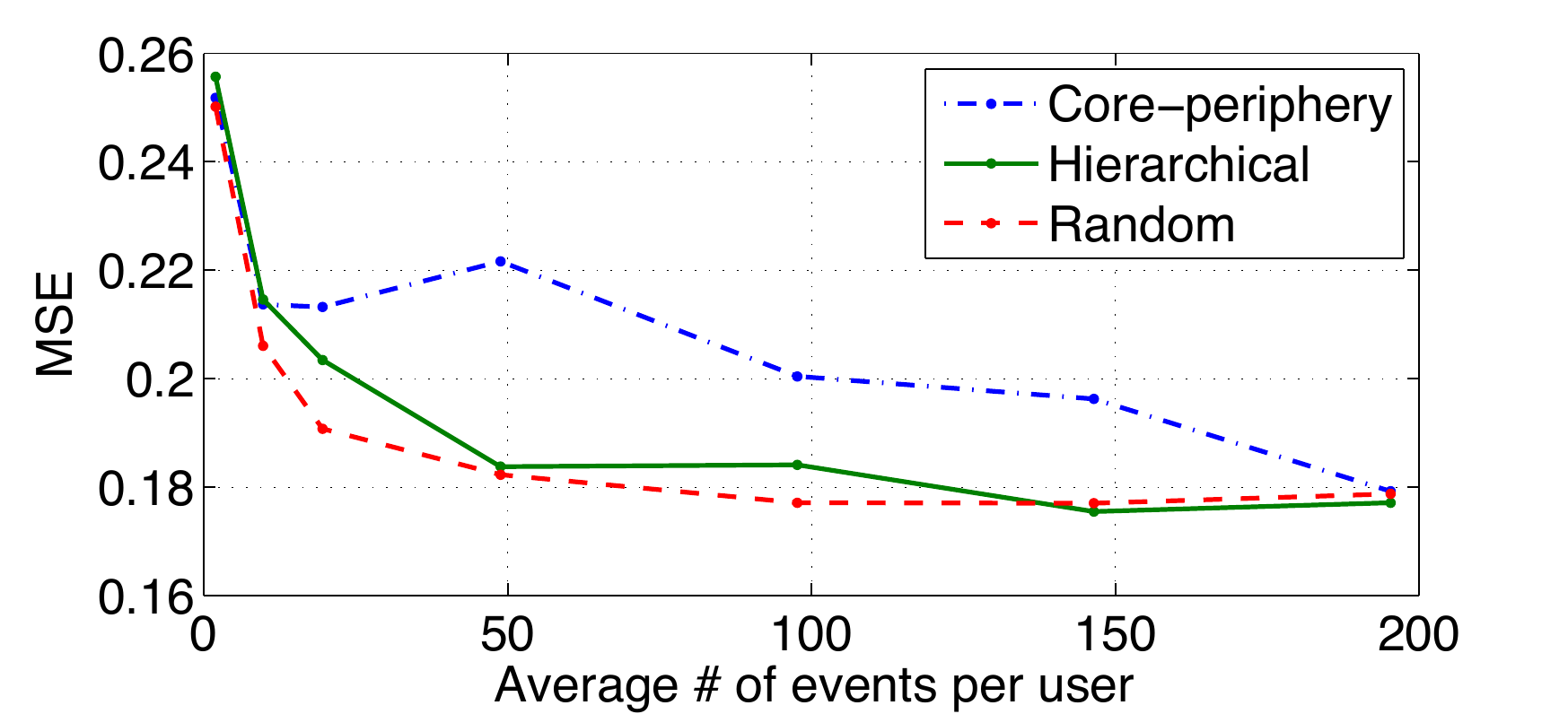}
%\vspace{-2mm}
\caption{Performance of our model estimation method in three Kronecker networks. We set the num\-ber of nodes in the three networks to $512$ users and the number of edges 
to $2{,}040$ (Core-Periphery), $4{,}608$ (Hierarchical) and $7{,}669$ (Random).}
%\vspace{-2mm}
\label{fig:MSE}
\end{figure}
In this section, we validate our model using synthetic networks and product use event data.
In particular, we will demonstrate that our model estimation method can accurately recover the true model parameters from his\-to\-ri\-cal event data using synthetic data. 
\begin{figure}[t]
\centering
\subfigure[Events]{\includegraphics[width=.36\textwidth]{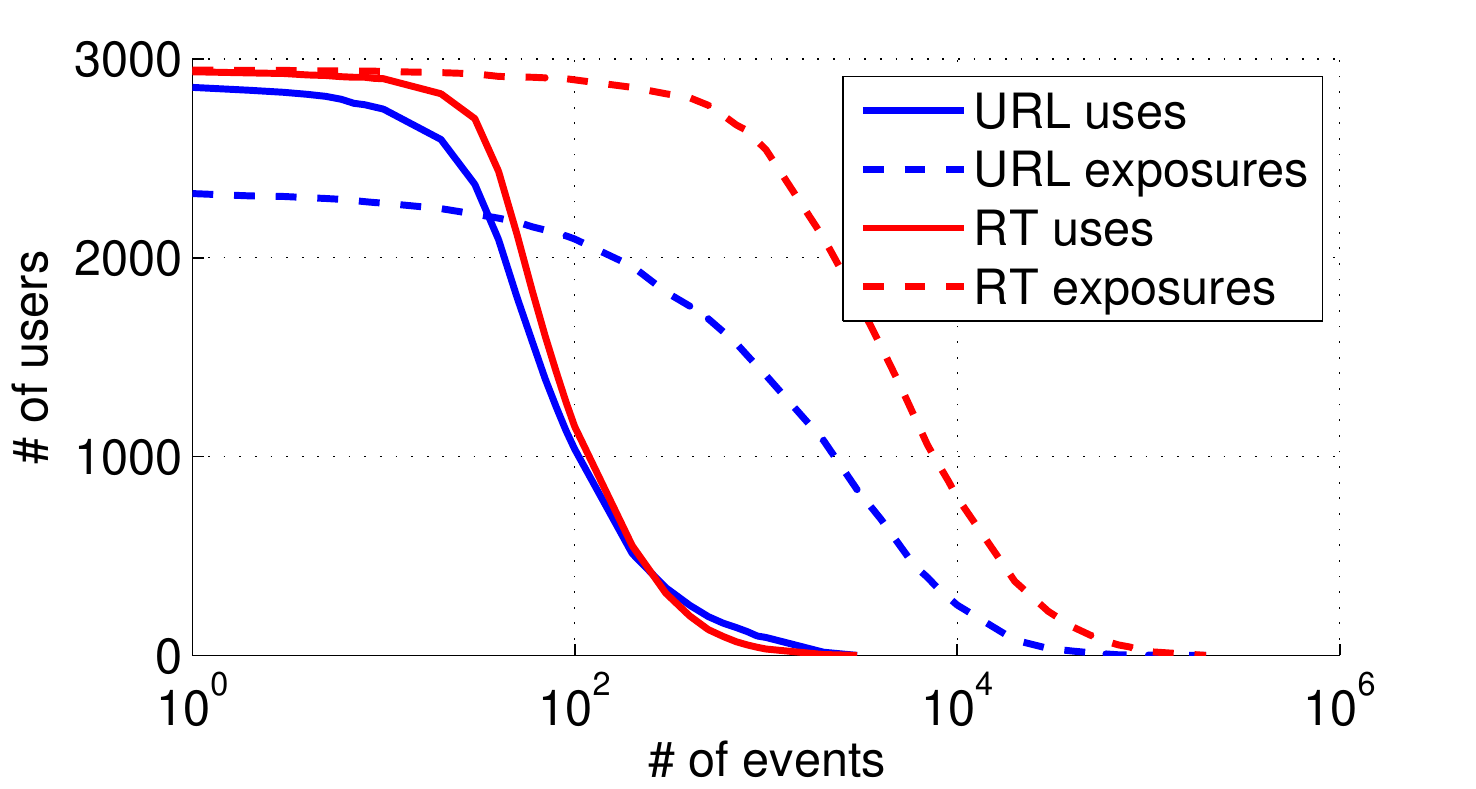}\label{fig:analyticsEvents}} \\
%\vspace{-2mm}
\subfigure[Products]{\includegraphics[width=.36\textwidth]{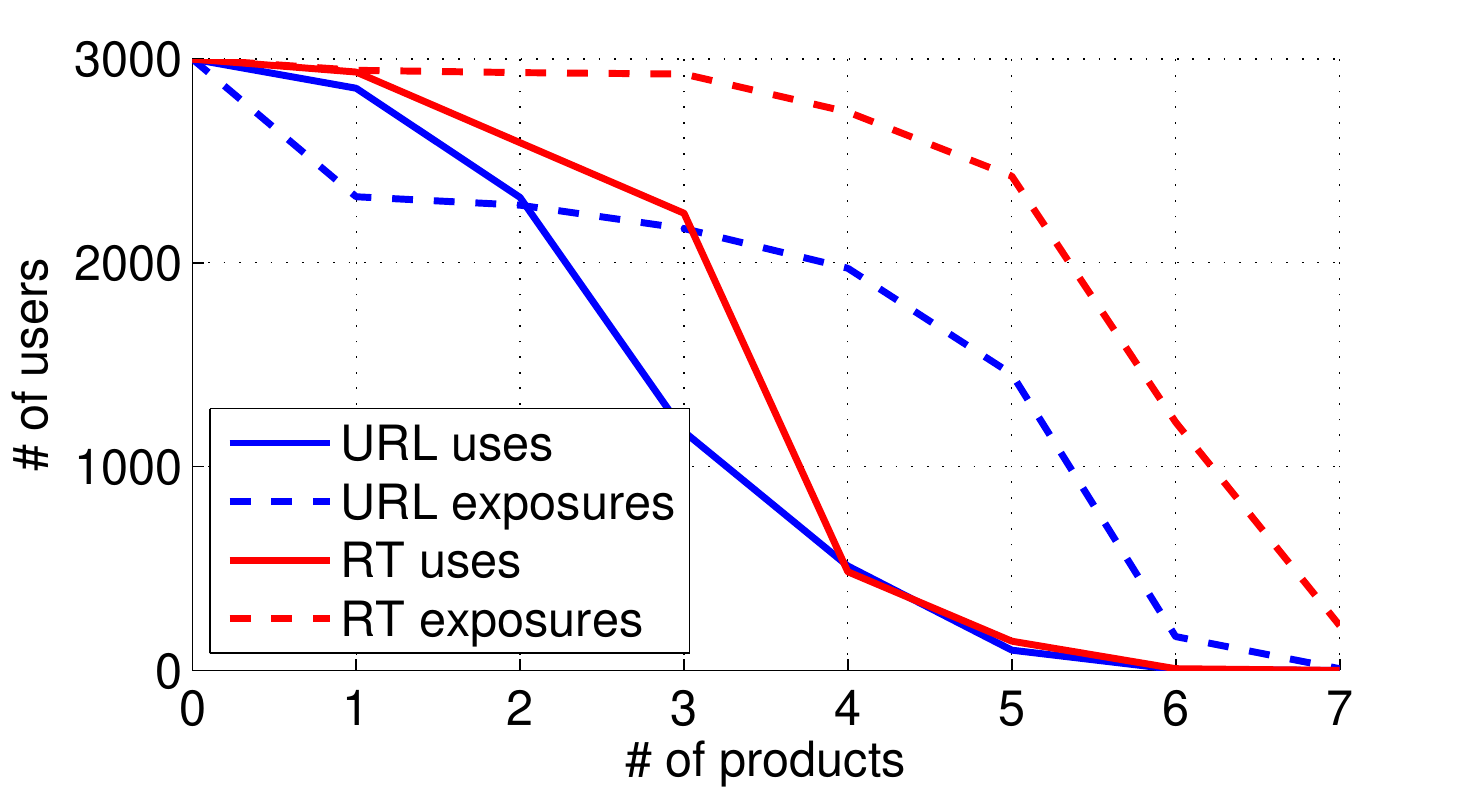}\label{fig:analyticsProducts}} 
%\vspace{-2mm}
\caption{Product use event statistics. Panel (a) shows the number of users that generated (solid lines) or got exposed to (dashed lines) at least $x$ use events during the considered 4-month period. 
Panel (b) shows the number of users that used (solid lines) or got exposed to (dashed lines) at least $x$ different products during the considered 4-month period.}
%\vspace{-2mm}
\label{fig:analytics}
\end{figure}

\subsection{Experimental setup}
We generate three types of synthetic networks using a well-known model of directed social networks, the Kronecker graph model~\cite{leskovec2010kronecker}:
%and the Forest Fire (scale free) model~\cite{barabasi99emergence}. 
(i) {\em core-periphery} networks~\cite{jure08ncp} (parameter matrix: $[0.9, 0.5; 0.5, 0.3]$), (ii) {\em hierarchical} networks~\cite{clauset08hierarchical} ($[0.9, 0.1; 0.1, 0.9]$), and 
(iii) {\em random} networks ($[0.5, 0.5; 0.5, 0.5]$).
%and a Forest Fire network with forward burning probability $0.15$ and backward burning probability  $0.1$. 
%
% \manuel{We mention number of nodes in the caption of the figure}
% We set the number of nodes in the three networks to $512$ users and the number of edges to $2{,}040$, $4{,}608$ and $7{,}669$, respectively. 
%The obtained number of edges in the Forest Fire network is $10{,}639$.
%
Then, for each network, we assume there are two competing products, and set the model parameters for each product and node in the networks as follows.  
First, we draw $\{a^{u}_{pp}\}$ and $\{b^{u}_{pp}\}$ from $U(0, 1)$ and $\{a^{u}_{lp}\}_{l\neq p}$ and $\{b^{u}_{lp}\}_{l\neq p}$ from $U(-1, 1)$. 
Then, we allow for a small set of users to have a baseline parameter greater than zero, and draw their baseline parameters from the uniform 
distribution $U(0, 1)$. 
% In this way, we account for the fact that only a few nodes acquire products spontaneously.
%
%
%Here, the baseline rate account for spontaneous use of a product due to exogenous factors to the network , and the single product parameters and the cross-product parameters account for 
%competition (and more rarely cooperation) between different products by network exposure.
%
Finally, for each network, we generate and record a set of $100{,}000$ events using our efficient sampling method, summarized in Algorithm~\ref{alg:simulation}. 
%
% \manuel{We already mention this earlier in the paper}
% Note that, since we allow 
% negative values for $\{a^{u}_{lp}\}_{l\neq p}$ and $\{b^{u}_{lp}\}_{l\neq p}$, an intensity function can be ne\-gative at some $t$ and thus become ill defined. As discussed in Section~\ref{}, 
% we simply trim it to zero, and assume that a node cannot draw samples as long as it is zero, as proposed previously~\cite{ogata1998linear}. 
%
%The procedure to generate samples from the multivariate Hawkes process is summarized in Algorithm~\ref{alg1}, where we assume a $|\Vcal|\times |\Pcal|$-dimensional 
% Hawkes process, in which each intensity function correspond to a user-product pair and is given by Eq.~\eqref{eq:hazard-general}.
%
%
Now, given the times when each user used any of the two products, our goal is to find the true model parameters $\boldsymbol{\mu}^{u}$, $\mathbf{A}^u$, and $\mathbf{B}^u$ for each 
user $u$, by solving the regularized maximum likelihood estimation (MLE) problem defined in Eq.~\ref{eq:opt-problem}.

\subsection{Model Estimation}
We evaluate our model estimation method by comparing the inferred and true parameters in terms of the mean squared error (MSE), $E\left[\left(x-\hat{x}\right)^2\right]$, where $x$ is
the true parameter and $\hat{x}$ is the estimated parameter.
%
% In particular, for each of the three networks, we compute the MSE across users, products and all model parameters.
% , i.e., the baseline $\mu_p$, the product parameters $\{a^u_{pp}\}_{u \in \Vcal}$ and $\{b^u_{pp}\}_{u \in \Vcal}$, and the cross-product parameters, $\{a_{lp}\}_{u \in \Vcal,l \neq p}$ and $\{b_{lp}\}_{u \in \Vcal,l \neq p}$. 
%
Figure~\ref{fig:MSE} shows the MSE against average number of events per user for three Kronecker networks, where we set the regularization parameter $\beta=10$ and the parameter of the
trigge\-ring kernels $\omega=1$. 
We observe that the MSE decreases as the average number of events per user increases, reaching values below $0.18$ for the three networks once we observe $\sim$$200$ product
use events per user in average.
It is important to note that even though the networks have very different global network structures, the performance of our estimation method is remarkably stable and
does not seem to depend on the structure of the network.
%
%In particular, for each product, we compute the MSE across users on the baseline, $\mu_p$, on the product parameters, $a_{pp}$ and $b_{pp}$, and on the cross-product parameters, 
%$\{a_{lp}\}_{l \neq p}$. \manuel{Maybe these are too many lines to plot for three networks?}.

%\begin{itemize}
%\item We estimate the parameters of the model from the events you just generated. I would say we measure average 
%mean square error on the parameters (across nodes) against number of either total number of exposures to events per node, or total number of infections per node, or something on those lines.
%
%\item During the simulation, if a hazard is ill defined at some point, I would say we simply trim it to zero, and we assume that node cannot draw samples as long as it is 0 (the inverse transformation
%would give an infinite time). What do you think? Previous work~\cite{ogata1998linear} does that.
%
%\item In the end, we would simply have a couple of figures that shows number of exposures, infections or in general amount of generated data, vs MSE for the three model cases and 2-3 synthetic
%networks.
%\end{itemize}
% 

\section{Experiments on real data} \label{sec:real-experiments}

In this section, we validate our model using real-world networks and product use event data.
We first show that our model can accurately predict product use events in real-world data gathered from Twitter~\cite{cha2010measuring}, significantly 
outperforming a state of the art method~\cite{anderson2014dynamics} and two baselines.
We then use our model to derive insights into some of the factors driving product adoption and frequency of use, \ie, social influence, recency 
and product competition\-.

\subsection{Experimental Setup} 
We use data gathered from Twitter as reported in previous work~\cite{cha2010measuring}, which comprises of $1.7$ \mbox{billion} public tweets posted by $52$ million users during
a three year period, from March 2006 to September 2009. 
Importantly, this dataset is complete, meaning that it contains all public tweets posted before the end of September 2009.
Based on this raw data, we build two dataset of pro\-duct use events.
In the first dataset, we track every use event of the seven most popular (used) url shor\-te\-ners~\cite{antoniades2011we}: Bitly, TinyURL, Isgd, TwURL, SnURL, Doiop and Eweri. 
Url shorteners exis\-ted before Twitter, \mbox{however}, by constraining the number of characters per message, Twitter increased the proliferation of new url shorteners as well as their usage. 
In the second dataset, we track every use event of the seven most popular retweet conventions\footnote{The way Twitter users indicated back in 2009 that a tweet was being retweeted 
(or forwarded). Here, we think of diffe\-rent retweet conventions as competing products.}~\cite{kooti2012emergence}: RT, via, HT, retweet, retweeting, the symbol \includegraphics[scale=.5]{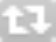} and R/T.
Retweet conventions emerged organically during the first few years of Twitter and until November 2009, when Twitter rolled out an official built-in retweet button.
In both cases, the functionality (or quality) of each competing product is si\-mi\-lar, therefore, one may expect social influence, recency and competition factors to play a more significant role than product 
qua\-li\-ty~\cite{anderson2014dynamics}. This allows us to directly observe and quantify the impact of each of these factors in the behavior pattern of each user in the network. 
It would be also interesting to consider competing products with clearly different quality and augment our model to incorporate product quality as a covariate in the product intensity defined 
by Eq.~\ref{eq:hazard-general}.
\begin{table}[t]
\centering
\renewcommand{\tabcolsep}{8pt}
\renewcommand{\tabcolsep}{3pt}
\subfigure[Average prediction probability across all users.]{\begin{tabular}{|c|c|c|c|c|} \hline
Data &  Our model  & Poisson & Weibull  & Recency~\cite{anderson2014dynamics} \\ \hline\hline
URL & $\mathbf{0.83}$ & $ 0.64$ & $0.45$ & $0.78$ \\ \hline
RT & $\mathbf{0.82}$ & $0.65$  & $0.78$ & $0.81$\\ \hline
\end{tabular}} \\
\vspace{2mm}
\subfigure[Percentage of users for which each model achieves the maximum prediction pro\-ba\-bi\-li\-ty.]{\begin{tabular}{|c|c|c|c|c|} \hline
Data &  Our model & Poisson & Weibull  & Recency~\cite{anderson2014dynamics} \\ \hline\hline
URL &  $\mathbf{81.6}$\% & $32.8$\% & $58.2$\% & $41.3$\%\\ \hline
RT & $\mathbf{89.1}$\%  & $45.2$\% &$76.3$\% & $57.3$\% \\ \hline
\end{tabular}}
\vspace{5mm}
\caption{Prediction Probability.}\label{tab:Prediction}
%\vspace{-3mm}
\end{table}

% \xhdr{Experimental setup}
%
% Our goal is to estimate each user'{}s susceptibility to adopt (and repeatedly use) every product or social convention.
% ; in other words, find the optimal parameters $\boldsymbol{\mu}^u$, $\mathbf{A}^u$ and $\mathbf{B}^u$ for each user $u$. % , by solving the maximum likelihood problem defined in Eq.~\ref{eq:opt-problem}.
%
%

For each of the two datasets, we experiment with four consecutive (arbitrary) months of product event data\footnote{January 15 to May 15 for url shorteners; February 1 to May 31 for retweet conventions.}, 
where we filter out users that join Twitter less than one month before the start of the period, do not use any product after the period, or generate less than $100$ product use events during the whole three 
years of data, and focus on the 3,000 most active users (within the period) and their neighbors.
We build each user'{}s neighborhood using the interactions via @-messages; we create a directed edge $(i, j)$ as soon as user $j$ mentions user $i$ in a tweet, since this 
provides evidence that user $j$ is paying attention to user $i$~\cite{huberman2009}.
Then, we assume node $i$ got exposed only to product use events from node $j$ that occurred later than this first mention, as argued in previous work~\cite{romero11twitter}. 
Here, one could think of building each user'{}s neighborhood by simply considering all the users she follows, however, this would require obtaining the times in which each user starts following others, 
and these times are currently not provided by the Twitter public API~\cite{MyeLes14}.
In the remainder of the section, for each dataset, we employ the first three months of data as training set, and the last month as test set. 

Figure~\ref{fig:analytics} provides general statistics on the number of events users generate and are exposed to through her neighbors in both datasets.
We find that $\sim35\%$ of the url shortener users and $\sim75\%$ of the retweet convention users use three or more products. Moreover, retweet conventions users are exposed to 
a larger number of product use events and products than url shor\-te\-ners users.
%

% We would like to point out that our dataset also includes a snapshot of the followers/followees network by September 2009. However, we followed the above mentioned procedure because 
% the snapshot of the network does not contain the edge creation times and therefore, we cannot tell when a user $j$ started following a user $i$ from the snapshot of the network itself.
%
%
\begin{table}[t]
\centering
\renewcommand{\tabcolsep}{8pt}
\subfigure[Average test log-likelihood per event a\-cross all users.]{
\begin{tabular}{|c|c|c|c|} \hline
Data &  Our model  & Poisson & Weibull \\ \hline\hline
URL & $\mathbf{-1.09}$ & $ -3.28$ & $-1.18$\\ \hline
RT & $\mathbf{2.11}$ & $ -3.96$ & $-1.76$\\ \hline
\end{tabular}} \\
\vspace{2mm}
\subfigure[Percentage of users for which each model achieves the maximum average test log-likelihood per event.] {
\begin{tabular}{|c|c|c|c|} \hline
Data &  Our model & Poisson & Weibull \\ \hline\hline
URL &  $\mathbf{80.2}${\bf \%}& $4.1$\% & $15.7$\%\\ \hline
RT & $\mathbf{94}${\bf \%}  & $0$\% & $6$\% \\ \hline
\end{tabular}}
\vspace{5mm}
\caption{Average test log-likelihood.}\label{tab:Lik}
%\vspace{-5mm}
\end{table}

\subsection{Implementation and scalability}
We developed an efficient distributed implementation of our model estimation method, described in Section~\ref{sec:estimation}, and deployed it in a cluster with 
$1{,}000$ CPU cores. This allowed us to efficiently model thousands of users.
Although we considered five url shor\-te\-ners and retweet conventions, in practice, due to the cross validation step of the regularization parameter $\beta$ and the triggering
kernel parameter $\omega$, we experimented with $38$ different configurations per product and user. 
Therefore, since we mo\-de\-led $3{,}000$ users for both url shortener and retweet conventions, the analysis involved estimating over $50$ million model parameters by solving 
$1{,}140{,}000$ ($3{,}000 \times 5 \times 38 \times 2$) convex optimization problems, in which the number of events in the objective function spans from $0$ to hundreds 
of events.
Remarkably, for each optimization problem, we can precompute the sums of triggering kernels and integrals of triggering kernels in a time linear in the number of events a 
user is exposed to, as discussed in Section~\ref{sec:estimation}. 
As a consequence, the overall fitting procedure for a very active user never took more than a few seconds and, for a less active user, it often took only a few milliseconds.

\subsection{Product use event prediction}
We eva\-luate the perfor\-mance of our model in comparison with one state of the art method~\cite{anderson2014dynamics} (`Recency'),  which is the only approach in the literature that deal with recurrent usages of products and not only for single adoptions, and therefore allows for fair comparison with ours.
We also compare our model to two baselines, a memoryless Poisson process (`Poisson')~\cite{Kingman1992} and 
a Weibull renewal process (`Weibull')~\cite{barabasi05human}.
The Recency model is a discrete time model that considers a product from $t$ events ago to be reused with a probability proportional to a function of $t$.
Poisson and Weibull are continuous time models that consider, respectively, a constant (memory-less) intensity, as defined in Eq.~\ref{eq:poisson}, and an intensity that depends on the last event, as defined in Eq.~\ref{eq:weibull}.
For each user, we fit all models using maximum likelihood estimation. 
In our model, we set the regularization parameter $\beta$ and the decay function parameter $\omega$ via cross-validation. 
% In the cross-validation step, we select  the parameters $\beta$ and $\omega$ that maximize the likelihood per user $u$ on the last month of the training set data, which has not previously used to solve the optimization problem.
%
% Moreover, we force the cross-product parameters $a^{u}_{lp}$ and $b^{u}_{lp}$ for all $l\neq p$ to be equal or smaller than zero, since, in practice, 
% this provides more stable solutions. 
%
% For the memoryless Poisson and the Weibull renewal models, we fit one model per user $u$ and product $p$ using maximum likelihood estimation. 
%
In the Recency model, we consider a memory equal to five events.

Then, we use three different evaluation measures to compare our model to the other three models: 
%
% \vspace{-1mm}
\begin{enumerate}
%\denselist
%
\item[I.] {\bf Prediction Probability:} For each user and (trained) model, we predict the identity of the product associated to each true event in her test set by evaluating and ranking the likelihoods (probability for the Recency model) asso\-cia\-ted to each product at 
the test event and taking the top-1 product. 
The prediction probability is then defined as the probability that the true and the predicted identity for each product use test event coincide. 
This evaluation measure has been used previously by the authors of the Recency method~\cite{anderson2014dynamics}.

\item[II.] {\bf Average Test Log-likelihood:} For each user and continuous-time (trained) model, we compute the average log-like\-li\-hood per event in the test set.
This allows us to assess how well each continuous-time model fits the true times of each test event -- in other words, its goodness of fit.
Note that we compute the log-likelihood over the test data, not the training data, to measure how \emph{well} each model generalizes 
to the data.
Here, we cannot compute the average test log-likelihood for the discrete-time Recency model, since it does not model the actual time of each product 
use event.

\item[III.] {\bf AIC:} For each user and continuous-time (trained) mo\-del, we compute the AIC~\cite{AIC} (Akaike information criterion). The AIC is given by $AIC = 2N_p - 2\mathcal{L}$, 
where $N_p$ is the number of model parameters and $\mathcal{L}$ is the average log-likelihood evaluated in the training set. 
AIC quantifies the trade-off between the goodness of fit for the true times of each training event and the complexity of the model and thus has been typically used for model selection. 
The preferred model is the one with the minimum AIC value.
As with the average test log-likelihood, we cannot compute the AIC for the discrete-time Recency model, since it does not model the actual time of each 
product use event.
\end{enumerate}
\begin{table}[t]
\centering
\renewcommand{\tabcolsep}{8pt}
\subfigure[Average AIC across all users. Lower is be\-tter.]{
\begin{tabular}{|c|c|c|c|} \hline
Data &  Our model  & Poisson & Weibull \\ \hline\hline
URL &  $\mathbf{-430.4}$ & $-6.4$ & $-66.6$\\ \hline
RT & $\mathbf{-44.6}$ & $ 25.5$ & $24.5$\\ \hline
\end{tabular}
}  \\
\vspace{2mm}
\subfigure[Percentage of users for which each model achieves the maximum AIC.]{
\begin{tabular}{|c|c|c|c|} \hline
Data &  Our model & Poisson & Weibull \\ \hline\hline
URL &  $\mathbf{65}${\bf \%}  & $2.5$\% & $32.5$\%\\ \hline
RT & $\mathbf{66.8}${\bf \%}  & $6.1$\% & $27.1$\% \\ \hline
\end{tabular}
}
\vspace{5mm}
\caption{Akaike information criterion (AIC).}\label{tab:AIC}
%\vspace{-3mm}
\end{table}

\begin{figure}[t]
\centering
\subfigure[Twitter user'{}s url shortening service usage pattern]{\includegraphics[width=0.45\textwidth]{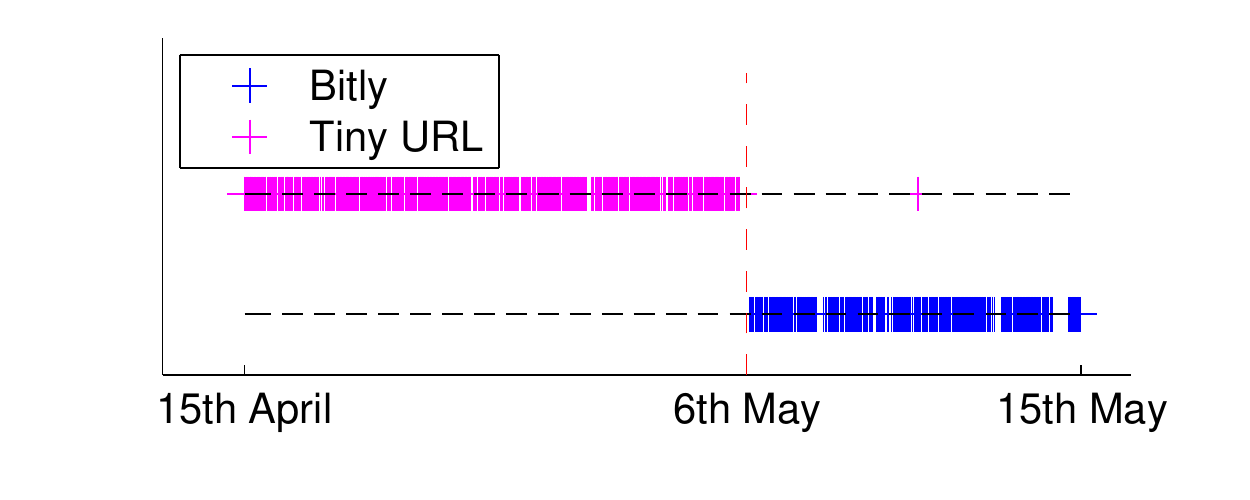} \label{fig:EvEx}} \\
\subfigure[Average test log-likelihood per event and user]{\includegraphics[width=0.45\textwidth]{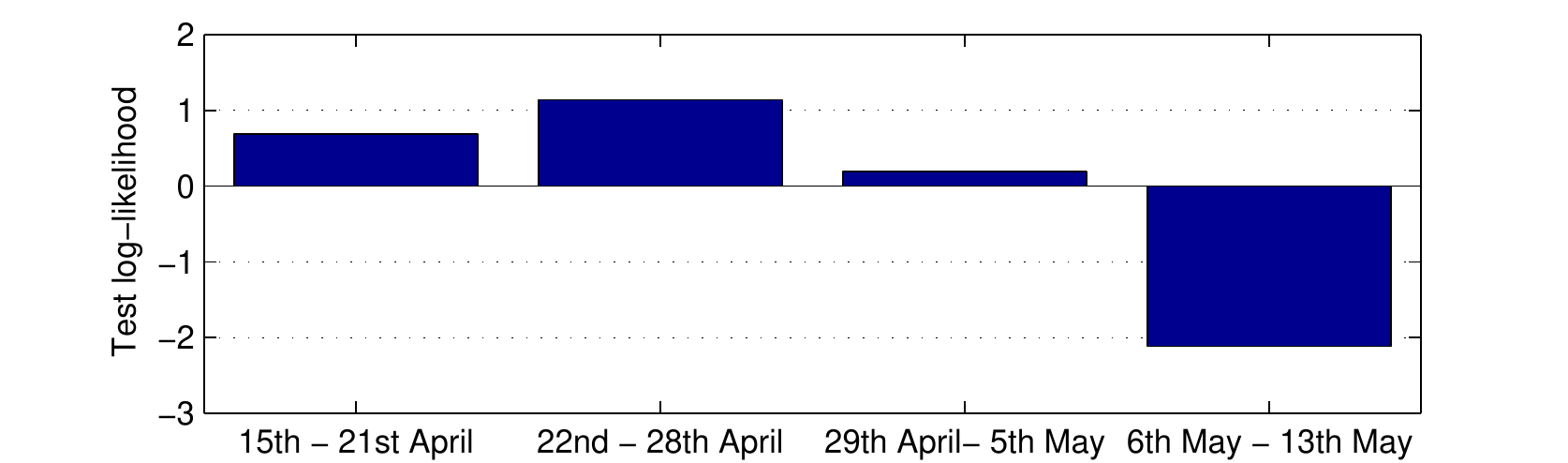} \label{fig:URLdefault}}
% \vspace{-3mm}
\caption{An example of external intervention. On May 6th, 2009, Twitter changed the default url shortener from `tinyurl.com' to `bit.ly'.} \label{fig:ExternalIntervention}
% \vspace{-4mm}
\end{figure}

Table~\ref{tab:Prediction}(a) shows the average production prediction pro\-ba\-bi\-li\-ty across all users. Our model outperforms the Recency model and both baselines for
both url shorteners and retweet conventions. 
At first, one may think that the improvement with respect to the Recency model is small\footnote{The difference is statistically significant ($p=$).}, however, Table~\ref{tab:Prediction}(b) 
demonstrates that our model consistently achieves the highest prediction pro\-ba\-bi\-li\-ty for each individual user by showing the percentage of users for which each model provides the 
maximum product prediction probability across all models. 
In terms of individual users, our model beats the second best, the Weibull model, by more than $20$\% for url shorteners and more than $10$\% for retweet 
conventions while the Recency and the Poisson models achieve an underwhelming performance.
Here, note that the percentages sum up more than 100\% because whenever we encounter a user that generate a few use events during the test period, several models 
often provide the same maximum prediction probability.

Table~\ref{tab:Lik}(a) shows the average test log-likelihood per event across users. Our model outperforms both continuous-time baseline models for url shorteners and 
retweet conventions.
Moreover, Table~\ref{tab:Lik}(b) demonstrate that our model consistently achieves the highest average log-likelihood across the continuous-time models for each individual
user by showing the percentage of users for which each model provides the maximum test log-likelihood. In particular, our model is the best fit for as much as $80.2\%$ and 
$92\%$ of url shortening services and the retweet conventions users, respectively.
Note that probability densities of continuous random variables can be larger than $1$, and thus test log-likelihoods be larger than $0$ (\eg, our model achieves an average
test log-likelihood value of $2.11$ for retweet conventions).

Table~\ref{tab:AIC}(a) shows the average AIC across users and Table~\ref{tab:AIC}(b) the percentage of users for which each model provides the minimum AIC. Our model
beats both continuous-time baseline models in more than $65\%$ of the users.
Hence, we can conclude that our model is not only the model with the best performance in terms of product prediction pro\-ba\-bi\-li\-ty and test log-likelihood, but it also provides 
the best trade-off between goodness of fit and model complexity.

\subsection{External interventions} 
We have shown the effectiveness of our model at predicting the adoption and frequency of use of competing products. However, what if there is a \emph{sudden} exogenous change, 
which cannot be explained through social influence, recency and product competition?
For example, on May 6th, 2009, Twitter changed the default url shortener from `tinyurl.com' to `bit.ly', trigge\-ring a significant increase (decrease) on the amount of new adopters of `bit.ly' 
(`tinyurl.com') shortly afterwards, as illus\-tra\-ted by Fig.~\ref{fig:EvEx}.
Can we at least detect such a change once it happened?
As Figure~\ref{fig:URLdefault} demonstrates for the above mention example, such change is easily identifiable by means of the average test log-likelihood per event across users, 
which drops dramatically.
%
% This does not account for \textit{sudden} changes in the dynamics of the network, and therefore, to avoid substantial deteriorations in the performance of the model, 
%
Therefore, once a change is detected, by means of the average log-likelihood per event, one may decide to re-train the model.

\subsection{Social influence, recency and competition}
We conclude our experiments by exploring the properties of the inferred users'{} parameters on real data both qualitatively and quantitatively. By doing so, we aim to gain insights into some of the factors driving 
adoption and frequency of use of competing products.
Here, we focus on the five most popular url shorteners and retweet conventions, since the parameters of the two less popular shorteners and retweet conventions are hardly ever different than zero.
%
%In this section, we focus on the (`Full') model defined by Eq.~\ref{eq:hazard-general}, since it is the more ge\-neral one. \mbox{Nevertheless}, the other more restrictive variations of the model also support similar conclusions. 
%
\begin{table}[t]
\centering
\renewcommand{\tabcolsep}{5pt}
\subfigure[URL shortening services]{
\begin{tabular}{|c|c|c|c|} \hline  
Product & $\mu^u_p>0$ & $a_{pp}^u>0$ & $b_{pp}^u>0$ \\ \hline\hline
Bitly & $100\% $ & $39.2\% $ & $39.4\% $ \\ \hline
Tiny URL & $100\% $ & $89.1\% $ & $82.3\% $ \\ \hline \hline
Isgd & $0\% $ & $16.3\% $ & $23.6\% $ \\ \hline
TwURL & $0\% $ & $11.2\% $ & $19.9\% $ \\ \hline
SnURL & $ 0\% $ & $ 2.9\% $ & $ 3.6\% $ \\ \hline
\end{tabular}} \\
\vspace{2mm}
\subfigure[Retweet conventions]{
\begin{tabular}{|c|c|c|c|} \hline 
Convention & $\mu^u_p>0$ & $a_{pp}^u>0$ & $b_{pp}^u>0$ \\ \hline\hline
RT & $20.5\%$ & $94.9\%$  & $89.2\%$\\ \hline
via & $15\%$ & $59.6\%$ & $66.5\%$  \\ \hline
HT & $25.4\%$ & $60.2\%$ & $67.2\%$\\ \hline \hline
retweet & $11.1\%$ & $4\%$  & $7.5\% $\\ \hline
retweeting & $3.9\%$ & $3.2\%$  & $ 4.8\%$\\ \hline
%Recycle symbol (\includegraphics[scale=.5]{recycle}) & $0.15\%$ & $0.07\%$ & $0.07\%$ \\ \hline
% R/T & $0.59\%$ & $0.26\%$ & $0.30\%$ \\ \hline
\end{tabular}
}
\vspace{5mm}
\caption{Percentage of users with parameters $\mu^u_p>0$, $a_{pp}^u>0$ and $a_{pp}^u>0$.}\label{table:muab}
% \vspace{-5mm}
\end{table}

% \manuel{Where: In Figure~\ref{fig:Cum}, we show the true and the predicted total number of test events for the full') model. }
%
We start by comparing spontaneous adoption (by means of $\mu_p^u$), recency ($a_{pp}^{u}$) and (positive) social influence ($b_{pp}^{u}$) on the adoption and frequency of use of a product. Table~\ref{table:muab} summarizes 
the results, by showing the percentage of users for which $\mu^u_p$, $a^{u}_{pp}$ and $b^{u}_{pp}$ are greater than zero for url shorteners and retweet conventions.
We find clear differences between url shorteners and retweet conventions, which we summarize next.
On the one hand, Table~\ref{table:muab}(a) indicates that, for the $3{,}000$ users under study, spontaneous adoption dominates over recency or (positive) social influence to explain the use of popular shorteners (`bitly' and `tinyurl'). 
In contrast, less popular shorteners (`isgd', `twurl' and `snurl') are never used by spontaneous adoption but first used by (positive) social influence and then re-used by recency.
On the other hand, Table~\ref{table:muab}(b) indicates that recency and (positive) influence dominate over spontaneous adoption to explain the use of popular retweet conventions (`RT', `via' and `HT'). In contrast, spontaneous 
adoption, social influence and recency have a comparable strength on the adoption and use of less popular retweet conventions (`retweet' and `retweeting').

Next, we investigate competition (or inhibition) between products. Tables~\ref{table:As} and~\ref{table:Bs} summarize the results, by showing\- the percentage of users for which $a^{u}_{lp}$ and $b^{u}_{lp}$ (for $l\neq p$) are smaller 
than zero. 
Here, each value can be viewed as a measure of the degree of inhibition that an event of the product in the row entails on the usage of the products in each column.
We find several interesting patterns, which we summarize next.
First, if we pay attention to the last rows of each table, which correspond to less popular products and conventions, we conclude that using one of these products or conventions
has a strong inhibitory effect on using more popular product or conventions.
%
% This ensures that for the loyal users to the less popular products, the probability of using them holds greater than zero although these users are exposed to a large number of events of the other 
% more popular products, provided that the users or their neighbors keep on using that product. 
%
Second, we find that `bit.ly' has a stronger inhibitory effect on `tinyurl.com' but `tinyurl.com' does not have almost any effect on `bit.ly'. 
This may indicate that even before the change of default service from `tinyurl.com' to `bit.ly' on May 6th, `bit.ly' was already a strong competitor to `tinyurl.com'. 
%This result and the change of default service might explain the fact that the number of uses of 'bit.ly' is greater that the number of uses of 'tinyurl.com' although the former was adopted in Twitter 
% for the first time almost two years later than 'tinyurl.com'.
%
Finally, if we compare the inhibitory effect of previous product events by a user herself ($a^{u}_{lp}$) and her neighbors ($b^{u}_{lp}$), we find qualitatively similar values.

% Finally, the 'recycle button' inhibits the uses of any other product for a large number of users, that means that the 'recycle button' \textit{steals} users to the rest conventions. \manuel{Therefore, we 
% could say that once the 'recycle button' is available in Twitter (i.e., after September 2008) most of the users adopt the more easiest action which consists in pressing 'recycle button' to retweet. check 
% this!! Can we say that nowadays these conventions have almost disappeared in Twitter because the users resort to the retweet button?}

%\xhdr{Comparison with baselines}
%\begin{table}[t]
%\small
%\centering
%\renewcommand{\tabcolsep}{5pt}
%\begin{tabular}{|c|c|c|c|} \hline
%Bins Size & Hawkes& Poisson & Lin. Regression\\ \hline
%24h & 30.33 & 35.18 & 34.49 \\ \hline
%6h & 30.19 & 36.24 & 33.56 \\ \hline
%\end{tabular}
%\caption{URL shortening services. Percentage of users for which each model is the best.}\label{table:URLcomp}
%\vspace{-2mm}
%\end{table}
%
%\begin{table}[t]
%\small
%\centering
%\renewcommand{\tabcolsep}{5pt}
%\begin{tabular}{|c|c|c|c|} \hline
%Bins Size & Hawkes& Poisson & Lin. Regression\\ \hline
%24h & 30.25 & 29.61 & 40.10 \\ \hline
%6h & 29.77 & 28.01 & 42.22 \\ \hline
%\end{tabular}
%\caption{Retweet conventions. Percentage of users for which each model is the best.}\label{table:RTcomp}
%\vspace{-2mm}
%\end{table}
%%URLs 

%
\begin{table}
%\begin{table}[t]
\centering
\renewcommand{\tabcolsep}{5pt}
\subfigure[URL shortening services]{
\begin{tabular}{|c|c|c|c|c|c|c|} \hline
\backslashbox{$l$}{$p$} & Bitly & Tiny URL & Isgd & TwURL & SnURL \\ \hline \hline
Bitly&\cellcolor[cmyk]{0,0.1,0.4,0}  &$58.8\% $ &$9.7\% $ &$7.2\% $ &$1.3\% $  \\ \hline
Tiny URL&$2.64\% $ &\cellcolor[cmyk]{0,0.1,0.4,0}  &$1.1\% $ &$0.5\% $ &$0.2\% $  \\ \hline
Isgd&$28.2\% $ &$80.8\% $ &\cellcolor[cmyk]{0,0.1,0.4,0}  &$9.2\% $ &$1.4\% $ \\ \hline
TwURL&$33\% $ &$81.8\% $ &$21.6\% $ &\cellcolor[cmyk]{0,0.1,0.4,0}  &$2.8\% $  \\ \hline
SnURL&$36.6\% $ &$87.9\% $ &$24.6\% $ &$13.8\% $ &\cellcolor[cmyk]{0,0.1,0.4,0}   \\ \hline
%Bitly&\cellcolor[cmyk]{0,0.1,0.4,0}  &$61.18\% $ &$10.22\% $ &$ 7.29\% $ &$ 1.47\% $ % &$ 0\% $ &$ 0.04\%$ 
%\\ \hline \hline
%Tiny URL&$ 1.59\% $ &\cellcolor[cmyk]{0,0.1,0.4,0}  &$ 1.01\% $ &$ 0.42\% $ &$ 0.21\% $ % &$ 0\% $ &$ 0\%$ 
%\\ \hline \hline
%Isgd&$29.31\% $ &$85.51\% $ &\cellcolor[cmyk]{0,0.1,0.4,0}  &$ 9.38\% $ &$ 1.47\% $ % &$ 0\% $ &$ 0.13\%$ 
%\\ \hline \hline
%TwURL&$35.01\% $ &$86.68\% $ &$23.07\% $ &\cellcolor[cmyk]{0,0.1,0.4,0}  &$ 3.02\% $ % &$ 0\% $ &$ 0.17\%$ 
%\\ \hline \hline
%SnURL&$38.78\% $ &$93.63\% $ &$26.01\% $ &$14.41\% $ &\cellcolor[cmyk]{0,0.1,0.4,0}  % &$ 0\% $ &$ 0.17\%$ 
%\\ \hline
% Doiop&$41.12\% $ &$96.65\% $ &$28.60\% $ &$15.03\% $ &$ 4.19\% $ &\cellcolor[cmyk]{0,0.1,0.4,0}  &$ 0.17\%$ \\ \hline \hline
% Eweri&$40.87\% $ &$96.44\% $ &$28.48\% $ &$14.99\% $ &$ 4.06\% $ &$ 0\% $ &\cellcolor[cmyk]{0,0.1,0.4,0} \\ \hline
\end{tabular}} \\
\vspace{2mm}
\subfigure[Retweet conventions]{
\begin{tabular}{|c|c|c|c|c|c|c|} \hline
\backslashbox{$l$}{$p$} &RT & via & HT & Retweet & Retweeting \\ \hline  \hline
RT&\cellcolor[cmyk]{0,0.1,0.4,0}  &$1.6\% $ &$1\% $ &$0.1\% $ &$0.1\% $  \\ \hline
via&$25.8\% $ &\cellcolor[cmyk]{0,0.1,0.4,0}  &$10.7\% $ &$3.3\% $ &$2.6\% $  \\ \hline
HT&$22.2\% $ &$12.9\% $ &\cellcolor[cmyk]{0,0.1,0.4,0}  &$2.5\% $ &$2.3\% $  \\ \hline
Retweet&$52\% $ &$45.2\% $ &$40\% $ &\cellcolor[cmyk]{0,0.1,0.4,0}  &$1.6\% $  \\ \hline
Retweeting&$54.3\% $ &$62\% $ &$62.5\% $ &$4.4\% $ & \cellcolor[cmyk]{0,0.1,0.4,0}   \\ \hline
%RT&\cellcolor[cmyk]{0,0.1,0.4,0}  &$0.81\% $ &$0.63\% $ &$0.07\% $ &$0.11\% $ % &$0\% $ &0\% 
%\\ \hline \hline
%via&$41.64\% $ &\cellcolor[cmyk]{0,0.1,0.4,0}  &$10.5\% $ &$3.4\% $ &$2.11\% $ % &$0.04\% $ &0.15\% 
%\\ \hline \hline
%HT&$15.72\% $ &$5.84\% $ &\cellcolor[cmyk]{0,0.1,0.4,0}  &$1.55\% $ &$2.11\% $ % &$0.04\% $ &0.07\% 
%\\ \hline \hline
%Retweet&$54.84\% $ &$40.83\% $ &$38.76\% $ &\cellcolor[cmyk]{0,0.1,0.4,0}  &$1.52\% $ % &$0.15\% $ &0.41\% 
%\\ \hline \hline
%Retweeting&$47.86\% $ &$38.98\% $ &$39.98\% $ &$5.21\% $ &\cellcolor[cmyk]{0,0.1,0.4,0}  % &$0.11\% $ &0.15\% 
%\\ \hline
% Recycle Symbol (\includegraphics[scale=.5]{recycle})&$61.06\% $ &$46.41\% $ &$47.86\% $ &$9.43\% $ &$4.18\% $ &\cellcolor[cmyk]{0,0.1,0.4,0}  &0.48\% \\ \hline \hline
% R/T&$49\% $ &$38.91\% $ &$25.78\% $ &$5.55\% $ &$2.55\% $ &$0.15\% $ &\cellcolor[cmyk]{0,0.1,0.4,0} \\ \hline
\end{tabular}
}
\vspace{5mm}
\caption{Percentage of users with parameters $a^{u}_{lp}<0$.}\label{table:As}
% \vspace{-2mm}
\end{table}

%\section{Dynamics of competition}
%\label{sec:dynamics}
%\input{040dynamics}

\section{Conclusions}
\label{sec:conclusions}
We have developed a probabilistic model, based on temporal point processes, for the adoption and frequency of use of competing products in online social networks. 
By modeling recency, social influence and product competition, our model is able to generate realistic collections of product use events, obeying several temporal
pa\-tterns observed in real online social activity, as well as predict product use events more accurately than alternatives.

Our work also opens many interesting venues for future work.
For example, we have modeled the decay of the influence of an event over time using exponential triggering kernels $g_{\omega}(\cdot)$ and considered linear intensities with constant 
baseline rates. However, this is implicitly considering one characteristic time scale, which may be unrealistic in some scenarios.
A natural follow-up to potentially improve the accuracy of our model would be considering more complex triggering kernels~\cite{ZhoZhaSon13b}, performing a non-parametric estimation 
of the kernels~\cite{bacry2012non}, assuming time-varying baselines rates~\cite{aalen2008survival} or extending the model to incorporate nonlinearities~\cite{bremaud1996stability}. % or additional factors beyond social influence, recency and competition.
We have considered considered products and contagions with similar functionalities (or qualities) in order to study the impact of social influence, recency and competition play in the users' decisions. 
However, as future work, it would be also interesting to consider competing products with clearly different quality and augment our model to explicitly incorporate product quality.
Moreover, we have built each user'{}s neighborhood using the interactions via @-messages and considered each neighbor to have the same influence on a user. 
It would be interesting to build each user'{}s neighborhood using the actual follower network and consider each neighbor to have different influence\footnote{This would require obtaining the times in which each 
user starts following others and, unfortunately, these times are currently not provided by the Twitter public API~\cite{MyeLes14}.}.
Finally, our experimental validation relies on data gathered exclusively from Twitter, it would be interesting to evaluate our model in other microblogging (Weibo, Tumblr) 
and social networking (Facebook, G+) sites.
\begin{table}[t]
\centering
\renewcommand{\tabcolsep}{5pt}
\subfigure[URL shortening services]{\begin{tabular}{|c|c|c|c|c|c|c|} \hline
\backslashbox{$l$}{$p$} &Bitly & Tiny URL & Isgd & TwURL & SnURL \\ \hline  \hline 
Bitly&\cellcolor[cmyk]{0,0.1,0.4,0}  &$30.6\% $ &$2.6\% $ &$4.6\% $ &$0.6\% $  \\ \hline
Tiny URL&$6.5\% $ &\cellcolor[cmyk]{0,0.1,0.4,0}  &$1.8\% $ &$4\% $ &$0.3\% $  \\ \hline
Isgd&$30.1\% $ &$86.4\% $ &\cellcolor[cmyk]{0,0.1,0.4,0}  &$7.2\% $ &$0.9\% $  \\ \hline
TwURL&$29.7\% $ &$82.6\% $ &$15.6\% $ &\cellcolor[cmyk]{0,0.1,0.4,0}  &$1.6\% $  \\ \hline
SnURL&$31.5\% $ &$81.9\% $ &$17.9\% $ &$10.5\% $ &\cellcolor[cmyk]{0,0.1,0.4,0}   \\ \hline
%Bitly&\cellcolor[cmyk]{0,0.1,0.4,0}  &$31.70\% $ &$2.76\% $ &$4.61\% $ &$0.67\% $ % &$0\% $ &$0\%$ 
%\\ \hline \hline 
%Tiny URL&$5.31\% $ &\cellcolor[cmyk]{0,0.1,0.4,0}  &$1.76\% $ &$3.94\% $ &$0.29\% $ % &$0\% $ &$0\%$  
%\\ \hline \hline 
%Isgd&$28.89\% $ &$89.15\% $ &\cellcolor[cmyk]{0,0.1,0.4,0}  &$6.83\% $ &$0.92\% $ % &$0\% $ &$0\%$  
%\\ \hline \hline 
%TwURL&$29.43\% $ &$85.09\% $ &$16.16\% $ &\cellcolor[cmyk]{0,0.1,0.4,0}  &$1.67\% $ % &$0\% $ &$0\%$ 
%\\ \hline \hline 
%SnURL&$32.16\% $ &$85.76\% $ &$18.46\% $ &$10.51\% $ &\cellcolor[cmyk]{0,0.1,0.4,0}  % &$0\% $ & $0.04\%$ 
%\\ \hline
% Doiop&$42.16\% $ &$97.78\% $ &$29.48\% $ &$15.32\% $ &$4.18\% $ &\cellcolor[cmyk]{0,0.1,0.4,0}  & $0.16\%$ \\ \hline \hline 
% Eweri&$39.11\% $ &$94.22\% $ &$27.17\% $ &$14.44\% $ &$3.89\% $ &$0\% $ &\cellcolor[cmyk]{0,0.1,0.4,0} \\ \hline
\end{tabular}} \\
\vspace{2mm}
\subfigure[Retweet convention]{
\begin{tabular}{|c|c|c|c|c|c|c|} \hline
\backslashbox{$l$}{$p$} &RT & via & HT & Retweet & Retweeting \\ \hline  \hline
RT&\cellcolor[cmyk]{0,0.1,0.4,0}  &$2.1\% $ &$1\% $ &$0.2\% $ &$0.1\% $  \\ \hline
via&$27.6\% $ &\cellcolor[cmyk]{0,0.1,0.4,0}  &$1.8\% $ &$0.6\% $ &$0.6\% $  \\ \hline
HT&$26.5\% $ &$1.3\% $ &\cellcolor[cmyk]{0,0.1,0.4,0}  &$0.5\% $ &$0.6\% $  \\ \hline
Retweet&$56.6\% $ &$47.2\% $ &$19.6\% $ &\cellcolor[cmyk]{0,0.1,0.4,0}  &$1.6\% $ \\ \hline
Retweeting&$64.3\% $ &$61.1\% $ &$55.5\% $ &$2.6\% $ &\cellcolor[cmyk]{0,0.1,0.4,0}  \\ \hline
%RT&\cellcolor[cmyk]{0,0.1,0.4,0}  &$0.7\% $ &$0.48\% $ &$0.15\% $ &$0.07\% $ % &$0\% $ &0\% 
%\\ \hline \hline
%via&$34.28\% $ &\cellcolor[cmyk]{0,0.1,0.4,0}  &$2\% $ &$0.55\% $ &$0.37\% $ % &$0\% $ &0.04\% 
%\\ \hline \hline
%HT&$8.95\% $ &$0.89\% $ &\cellcolor[cmyk]{0,0.1,0.4,0}  &$0.37\% $ &$0.41\% $ % &$0\% $ &0.04\% 
%\\ \hline \hline
%Retweet&$58.14\% $ &$39.42\% $ &$28.37\% $ &\cellcolor[cmyk]{0,0.1,0.4,0}  &$1.48\% $ % &$0.04\% $ &0.15\% 
%\\ \hline \hline
%Retweeting&$47.3\% $ &$38.46\% $ &$35.95\% $ &$2.66\% $ &\cellcolor[cmyk]{0,0.1,0.4,0}  % &$0.04\% $ &0.22\% 
%\\ \hline
% Recycle Symbol (\includegraphics[scale=.5]{recycle}) &$71.01\% $ &$62.02\% $ &$62.83\% $ &$10.24\% $ &$4.62\% $ &\cellcolor[cmyk]{0,0.1,0.4,0}  &0.44\% \\ \hline \hline
% R/T&$53.14\% $ &$42.42\% $ &$36.28\% $ &$5.99\% $ &$3.11\% $ &$0.15\% $ &\cellcolor[cmyk]{0,0.1,0.4,0} \\ \hline 
\end{tabular}}
\vspace{5mm}
\caption{Percentage of users with parameters $b^{u}_{lp}<0$.}\label{table:Bs}
%\vspace{-2mm}
\end{table}

\bibliographystyle{IEEEtran}
\bibliography{refs}

\end{document}